\documentclass[12pt]{article} 
\usepackage[dvips]{graphics}
\title{ Collapse and the Tritium Endpoint Pileup}  
\author{{\it Richard Shurtleff~}\thanks{affiliation and mailing 
address: Department of Mathematics and Applied Sciences, 
Wentworth Institute of Technology, 550 Huntington Avenue, 
Boston, MA, USA, ZIP 02115, telephone number: (617) 989-4338, fax 
number: (617) 989-4591, page: www.public.wit.edu/faculty/shurtleffr/, e-mail address: shurtleffr@wit.edu}} 
\date{January 23, 1999}
\begin{document} 
          
\maketitle               
			\begin{abstract} 

The beta-decay of a tritium nucleus produces an entangled quantum system, a beta electron, a helium nucleus, and an antineutrino.  For finite collapse times, the post-collapse beta electron energy can originate from a range of pre-collapse energies due to the uncertainty principle. Long collapse times give negligible uncertainty, so the pre-collapse spectrum must approach that of isolated nuclei. We calculate the post-collapse electron spectrum which shows a collapse-dependent pileup near the endpoint. Comparison with observation shows that a collapse time of 1 x $10^{-17}$~s explains the observed pileup. The collapse of the entangled quantum system must be triggered by the environment: most likely an atomic (molecular) electron initially bound to the atomic (molecular) tritium source or perhaps ambient gas molecules. Coincidentally, the 40 eV tritium atom-helium ion energy level shift is unobservably small for times shorter than the system collapse time. We conclude that an atomic (molecular) electron triggers the collapse once the 40 eV shift becomes detectible and the electron detects the helium nucleus. Thus collapse may explain the tritium endpoint pileup.

	[Version 1.0, `Collapse Accounts for More at the Endpoint', prepared for the Fall Meeting of the New England Section of the American Physical Society at the University of New Hampshire, Durham, NH. Version 1.1 posted post-meeting at \linebreak www.public.wit.edu/faculty/shurtleffr/. This is version 2.0.]

	PACS number(s):  14.60.Pq and 23.40.-s

			\end{abstract}
\pagebreak

\section{Introduction} 

	What happens at the endpoint of tritium $\beta$-decay? The tritium nucleus decays to a helium nucleus, electron, and antineutrino with energy $E_{0} \approx $ 18.6 keV,
	\begin{equation} 
{\mathrm{H}}^3 \rightarrow {\mathrm{He}}^3 + e + \bar{\nu} + E_{0}.
	\end{equation}
What is the chance that an electron is emitted with energy $E_{1}$ just a little less than the endpoint energy $E_{0}$? Given the kinematics of a three-body decay and assuming the neutrino mass and the nuclear recoil energy are negligible, one can show [1] that the probability $P(E_{1}) dE_{1}$ that an electron is emitted into a window $dE_{1}$ at energy $E_{1}$ is $P(E_{1})$ = $N p_{1}^{2} (E_{0} - E_{1})^2,$ for $0 \leq  E_{1} \leq E_{0}$ and $P(E_{1})$ = 0 for $E_{0} \leq$ $E_{1},$ where $p_{1}$ is the electron momentum and $N$ is a normalization constant: $N$ = $1/\int_{0}^{E_{0}} (E_{2}^2 + 2 m_{e} E_{2}) (E_{0} - E_{2})^2 dE_{2}.$ The formula follows from the phase space available to the decay particles and neglects the existence of excited daughter atomic states, Coulomb corrections, and other complications. To simplify the discussion, we ignore all complications except the ones we need to introduce.

	Just below the endpoint an approximation suffices,
	\begin{eqnarray} 
  P(E_{1})   \approx 2 N m_{e} E_{0} (E_{0} - E_{1})^2 ,
\end{eqnarray}
where $E_{1} \approx$ $E_{0}$ and $m_{e}$ is the energy equivalent of the electron mass, $m_{e}$ = $511$ keV.  Thus we have

	\underline{Assumption I}. The probability that an electron is emitted with energy $E_{1}$ near the endpoint is given by the probability $P$ in (2).

	The observed spectrum [2-5] is something else. `An anomalous pileup of events at the endpoint' says the footnote in the Particle Data Group listing [6] for the neutrino mass squared, a comment attributed to Stoeffl and Decman [2]. An antineutrino mass would remove events, dropping the rate below that expected with the probability $P$ in (2), so an antineutrino mass is not the complication that solves the problem. Thus we have

	\underline{Observation}. An anomalous pileup of events is observed at the endpoint of tritium decay.

	Suppose we agree that (2) is wrong at the spectrometer by Observation, while by Assumption I the formula is correct at the decay site. This leaves the possibility that complications occur in-flight from decay to spectrometer that explain the change in spectrum from emission to detection.

\pagebreak

\section{In-Flight Complications} 

	Upon decay, the three decay particles form an entangled quantum system that eventually collapses into what we can picture as three separate, independent particle states, see Fig.~1. Some such systems would collapse quicker than others; let $T$ be a typical time between emission and collapse.

\begin{figure}[h] 	
\vspace{0in}
\hspace{0.75in}\includegraphics[0,0][360,180]{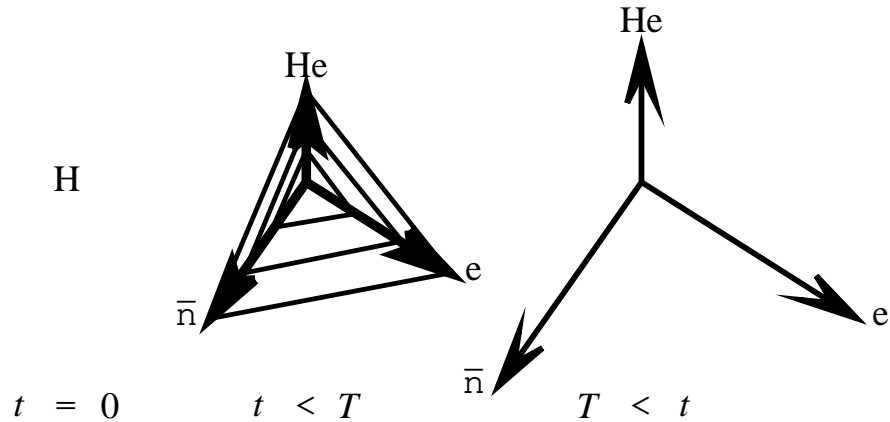}
\caption{Timeline. At time $t$ = 0 the tritium nucleus (H) undergoes $\beta$ decay. Until time $T$ the decay particles form an entangled quantum system. Some perturbation determines the state of the system at time $T$ so the system `collapses' into three single particle states. Long afterward, $T \ll t$, the electron reaches the spectrometer and its energy is measured.}
\end{figure}

	During time $t <$ $T$ the electron energy is uncertain within $\Delta E $ and $T \geq$ $\hbar/ (2 \Delta E)$, by one of the uncertainty principles [7]. Nothing can be done to find out what the electron energy is during this time without incurring an Immediate-Collapse penalty. One can certainly not use its eventual energy $E$ observed at the detector to infer the probability of emission $P$. If we knew $P$ at emission, then we would know the electron energy at emission and we would collapse the system immediately upon decay. 

	Now we have something to work with. The electron observed at the spectrometer with energy $E$ is an electron that until time $T$ had an energy $E_{1}$ somewhere in a range $\Delta E$ surrounding $E$. Thus we have

	\underline{Uncertainty Deduction}. An electron observed to have energy $E \leq$ $E_{0}$ could have been emitted with any energy $E_{1}$ in the range
	\begin{equation} 
E - \frac{\Delta E}{2} \leq E_{1} \leq E + \frac{\Delta E}{2}.
	\end{equation}
The spectrum $P$ in (2) is expected when $\Delta E$ is negligible and the typical collapse time $T$ is very long; $P$ describes the decay of isolated tritium nuclei.

The energies in the range (3) have different decay probabilities. For simplicity, we assume that an electron detected with energy $E$ could have originated with equal likelihood from anywhere in the range (3) with no contributions from energies outside the range. If $E_{2}$ and $E_{3}$ are both in the interval (3), then $P(E_{2})$ and $P(E_{3})$ contribute with equal weight to the probability average. By conservation of energy, and no matter what probability average is obtained for an energy $E >$ $E_{0}$, all detected electrons must have energy $E \leq$ $E_{0}$.

	\underline{Assumption II}. The probability that an electron arrives at the spectrometer with energy $E$ is the average $\bar{P}(E)$ over the emission probabilities $P(E_{1})$ for electrons emitted with energies $E_{1}$ in the range (3). 

\begin{figure}[h] 	
\vspace{0in}
\hspace{0.75in}\includegraphics[0,0][360,180]{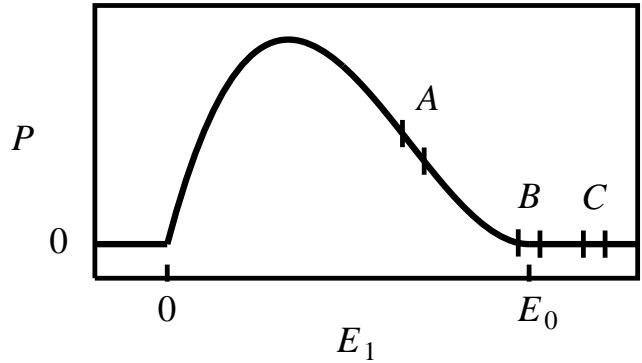}
\caption{The probability distribution of emitted electron energy. Three intervals $A$, $B$, and $C$ of width $\Delta E$ are shown. Interval $A$ is below the endpoint $E_{0}$, while $B$ contains the endpoint and $C$ is beyond the endpoint. We average the emission probability $\bar{P}$ over an energy interval to obtain the probability for the midpoint energy at the detector.}
\end{figure}

	Energy intervals like $A$ in Figs.~2 and 3 that do not contain the endpoint give an average emission into window $dE$ with probability $\bar{P}_{A}(E) dE$, where 
	\begin{eqnarray} 
\bar{P}_{A}(E) = \frac{1}{\Delta E} \int_{E-\Delta E/2}^{E+\Delta E/2} P(E_{1}) dE_{1} \approx P(E) + 2 N m_{e}E_{0} \frac{ \Delta E^2}{12},
	\end{eqnarray}
for $E+\Delta E/2 \leq$ $E_{0}$ and $E_{0} - E \ll$ $E_{0}$. 

\begin{figure}[h] 	
\vspace{0in}
\hspace{0.75in}\includegraphics[0,0][360,180]{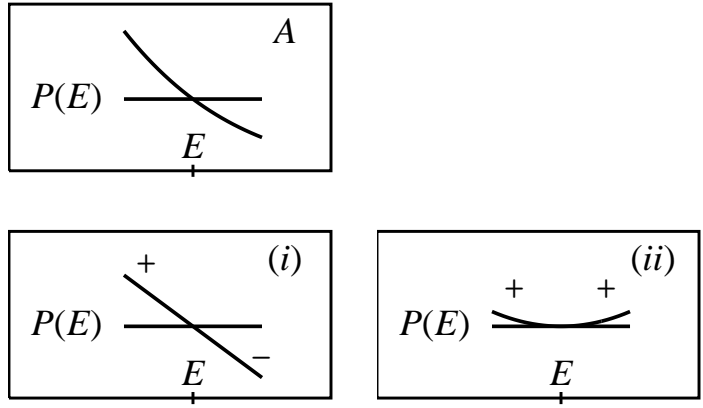}
\caption{The interval $A$ ranges from $E - \Delta E/2$ to $E + \Delta E/2.$ The linear part (i) and the quadratic part (ii) approximate the probability $P$ in the interval $A$. The linear part gives an average equal to the midpoint value, $\bar{P}_{(i)}$ = $P(E)$. The quadratic part gives extra events because the slope of the curve is increasing, thus $\bar{P}_{A} \approx$ $ \bar{P}_{(i)}$ + $\bar{P}_{(ii)} >$ $P(E)$, as in equation (4). [The change in slope has been exaggerated by plotting $P^5$.]}
\end{figure}


Intervals like $B$ in Fig.~2 that do contain the endpoint give 
	\begin{eqnarray} 
\bar{P}_{B}(E) = \frac{1}{\Delta E} \int_{E-\Delta E/2}^{E_{0}} P(E_{1}) dE_{1} 
	\end{eqnarray}
$$\approx 2 N m_{e}E_{0}\frac{(E_{0} - E+\Delta E/2)^3}{3 \Delta E},$$
for $E \leq$ $E_{0} \leq$ $E+\Delta E/2$. For intervals like $C$ in Fig.~2 with $E \geq$ $E_{0}$, the probability must vanish, $\bar{P}_{C}(E)$ = 0, to conserve energy.

	Let us consider the observations reported in Ref.~2. At $E$ = 18550 eV, Fig.~2 of Ref.~2 shows that $\bar{P}_{obs}$ = $1.2 P$. By (2), this means $\bar{P}_{obs}$ = $P$ + $2 N m_{e}E_{0} \times 70$ eV$^2 .$ By (4), $\bar{P}_{A}$ = $\bar{P}_{obs}$ when $\Delta E^{2} / 12$ = 70 eV$^2 .$ Thus $\Delta E$ = 30 eV.  ($\Delta E$ = 30 eV also satisfies $\bar{P}_{B}$ = $\bar{P}_{obs},$ but $E + \Delta E/2$ = 18565 eV $< E_{0},$ implying a type A interval; see Fig.~1.)

	\underline{Reconciliation Value} The electron spectrum adjusted for in-flight collapse agrees with the observed spectrum when the energy uncertainty is about $\Delta E$ = 30 eV.
In Fig.~4 we plot the spectrum (2) together with a spectrum with $\Delta E$ = 30 eV.

\begin{figure}[h] 	
\vspace{0in}
\hspace{0.75in}\includegraphics[0,0][360,180]{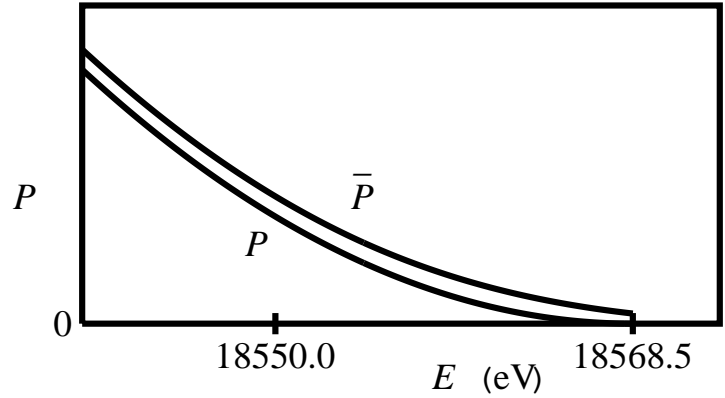}
\caption{The post-collapse spectrum $\bar{P}$ near the endpoint for the solution $\Delta E$ = 30 eV. At $E$ = 18550 eV, the number of observed events is about 20\% more than the spectrum $P$ for isolated tritium nuclei for which $\Delta E$ = 0. }
\end{figure}

\pagebreak

\section{Discussion} 

	The energy uncertainty gives a lower bound for $T$, 
	\begin{equation}
 T  \geq  \frac{\hbar}{2 \Delta E} \approx \frac{6.6 \times 10^{-16} \: {\mathrm{eV \cdot s}}}{2 \times 30 \: {\mathrm{eV}}} = 1.1 \times 10^{-17} \: {\mathrm{s}}.
	\end{equation}
This gives the collapse time for the three particle system that would provide an uncertainty of 30~eV in electron energy.

	An electron near the endpoint travels at a speed of $v$ = $ c \sqrt{1- m_{e}^2/(m_{e} + E_{0})^2}$ = 0.26$c$. In the average collapse time $T$ the electron travels a distance $x$ = $0.26 c T \geq$ $9 \times 10^{-10}$ m $\approx$ 9 atomic diameters and the antineutrino has traveled a distance of $cT \geq$ $33 \times 10^{-10}$ m $\approx$ 33 atomic diameters. The He nucleus recoils more than about $9 m_{e}/m_{{\mathrm{He}}}$  = $1.6 \times 10^{-3}$ of an atomic diameter. 

	Any of the three particles, electron, He nucleus, or antineutrino, could undergo the interaction or whatever it is that causes the collapse of the entire three particle entangled quantum system. At more than nine atomic diameters from the recoiling nucleus the electron could be interacting with the ambient gas molecules. Likewise, the antineutrino at more than 33 atomic diameters would be out amongst the gas molecules. 

	If the ambient gas forces collapse then varying the gas population should have an effect on the number of excess counts observed. A higher density makes for a shorter $T$ and a larger $\Delta E$. Doubling the gas density might double the pileup at 18550 eV.

	Alternatively, the recoiling He nucleus might be detected by the electron(s) originally bound to the tritium nucleus. The number of electrons bound to the tritium varies with the type of source: atomic tritium, molecular tritium, etc. If the bound electrons are involved in the collapse mechanism, then laser light of a frequency that is slightly more than the lowest resonance available to the original tritium sample might make the observed pileup a function of laser parameters.

	For simplicity consider an atomic tritium source. Immediately upon decay, the atomic electron remains in the tritium ground state $\psi_{{\mathrm{H}}}$. This state is a superposition of helium-3 states, with 70\% (= $\langle \psi_{{\mathrm{He}}} \mid \psi_{{\mathrm{H}}} \rangle^2$) of the electrons in the helium-3 ground state $\psi_{{\mathrm{He}}}.$ The change in the atomic electron's energy remains insignificant for a short time after the nucleus has decayed, by the uncertainty principle. 

	Assume that the atomic electron detects the changes in the nucleus shortly after the change in its own energy is detectible. Thus the collapse time $T$ would then be given by $T \geq$ $\hbar / (2 \delta E),$ where $\delta E$ is the change in the atomic electron's energy. Let us neglect the longer $T$s for other states and consider only the 70\% of atomic electrons that are in the helium-3 ground state after the nuclear decay. For these the energy change is $\delta E$ = $\mid E_{He} - E_{H} \mid $ = 40~eV and $T$ would be $T \geq  \hbar / (2 \delta E) \approx$ $ 0.8 \times 10^{-17}$~s. 

	The source in Ref.~2 is molecular tritium, but let us assume the results would be similar with atomic tritium. If 70\% of the tritium decays account for the observed 20\% excess at $E$ = 18550 eV, then the excess must be 30\% in the contributing population. Reworking the above calculation with the new excess gives $\Delta E^{2} / 12$ = $70 \times (30/20)$ eV$^2$ and $\Delta E$ is now 35 eV, $\Delta E$ = 35 eV. And $T$ decreases slightly to $T \geq  \hbar / (2 \Delta E) \approx$ $ 0.9 \times 10^{-17}$ s. The near coincidence of this result deduced from the observations, $T$ = $ 0.9 \times 10^{-17}$ s, and the collapse time deduced from the energy change of the atomic electrons, $T$ = $ 0.8 \times 10^{-17}$ s, implies that the endpoint pileup may be due to the detection of the tritium decays by the electrons bound to the decaying nuclei.

\pagebreak

\appendix

 \section{Problems} 

1. (i) Find the numerical value of the normalization constant $N$ in (2). Also (ii) find the energy $E_{{\mathrm{Max}}}$ in eV at the maximum probability and (iii) find $P_{{\mathrm{Max}}} dE$ for a $dE$ = 10 eV window centered on $E_{{\mathrm{Max}}}$. [(i) $9.8 \times 10^{-23}$ eV$^{-5}$. (ii) 6210 eV. (iii) 0.00096]

\noindent 2. A formula more accurate than the formula given in (2) is $P(E_{1})   \approx$ $2 N m_{e} E_{1} (E_{0} - E_{1})^2.$ Recalculate (4) and (5) with the more accurate formula and obtain the new uncertainty $\Delta E$ that gives $\bar{P}_{obs}(18550 \; {\mathrm{eV}})$ = $1.2 P(18550 \;{\mathrm{eV}})$.

\noindent 3. For a relativistic particle of mass $m$ and energy $E$ the momentum has magnitude $p$ satisfying $p^2$ = $E^2$ + $2 m E$. Show that this is true.

\noindent 4. For a given value of $\Delta E$ = 29 eV, plot the fractional change in the spectrum, $(\bar{P} - P)/P$ from 18340 to 18640 eV as in Fig.~2 of Ref.~2. On the same graph plot the result of averaging $\bar{P}$ with a Gaussian with a 10 eV width at half maximum to simulate a 10 eV spectrometer resolution.

\end{document}